# Beyond Superconductivity

**Jochen Mannhart**

Max Planck Institute for Solid State Research, 70569 Stuttgart, Germany

*Dedicated to Ted Geballe on the occasion of his 100$^{th}$ birthday.*

We present a novel device concept that utilizes the fascinating transition regime between quantum mechanics and classical physics. The devices operate by using a small number of individual quantum mechanical collapse events to interrupt the unitary evolution of quantum states represented by wave packets. Exceeding the constraints of the unitary evolution of quantum mechanics given by Schrödinger's equation and of classical Hamiltonian physics, these devices display a surprising behavior.

With his unique enthusiasm, profound understanding, and engaging excitement for venturesome science, Ted has inspired many for decades, certainly including myself. In the friendly sunshine of Santa Barbara's airport, for example, he suggested to me in his equally sunny style to continue pursuing his and Boris Moyzhes' ideas of electronic devices designed to convert heat extremely efficiently into electric power [1]. These devices work by utilizing electron motion in a plasma. Fortunately, we took his advice [2].

Now, on the joyful occasion of Ted's centennial birthday, I am happy to return the favor by presenting him and possible other interested readers with a comparable but even more daring device. Of course we state that it is even more efficient than his [2]. A secondary aspect which I expect will be of particular interest to him, is that such devices open new inroads to large, lossless, normal currents above room temperature – assuming that we are not thoroughly fooled by nature. This caveat obviously applies to my article as a whole.

Before embarking on a description of the device, I note that the specific example presented here is just one of a large family of devices that also include photonic systems [3]. This new concept exploits decoherence or collapse events intermingled with interference processes, as is only possible in the transition range between unitary quantum mechanics and classical physics.

The device presented here as example consists of an Aharonov–Bohm ring [4] made of a barely doped *n*-type semiconductor. The ring rests in a closed system in thermal equilibrium and is biased with a magnetic field (Fig. 1). Aharonov–Bohm rings have of course been explored intensively (see, e.g., [5,6]), but here we take a new perspective. We are interested in understanding how the rings behave if single inelastic scattering events occur during an electron's transit across the ring. For this, (1) we describe the electrons as wave packets, and not by infinitely extended planar waves as usual. (2) We consider that defects or phonons cause inelastic electron scattering also in the ring itself, and not exclusively in the contacts as assumed in the approach of Landauer and Büttiker [7]. (3) We take into account that scattering occurs as individual events that happen at a specific location and at a particular time. We find that their behavior is not adequately represented by the usual approach. It is generally assumed that the inelastic scattering merely results in noise and resistance, and washes out energy levels and interference effects. Finally, (4) we consider asymmetric rings; the two contacts are usually required to be exactly opposite. This perspective seems harmless and reasonable, yet it leads to surprising results.

First, we consider electron flow across such a ring in the absence of inelastic scattering. The path lengths along the left or right side of the ring and the magnetic flux reaching through the ring's hole are chosen such that electrons traveling from contact A to contact B undergo a phase change that differs between the two arms of the ring by $2n\pi$, whereas electrons moving from B to A acquire a phase difference of $2n\pi + \pi$, where $n$ is an integer [8,9].

As a result, electrons moving from A to B zip right through the ring, needing a time $\tau_{A \to B}$. Electrons traveling from B cannot leave the ring after their first traverse because they would then interfere destructively due to their $2n\pi + \pi$ phase differences. These electrons have to fly first forward through the ring, then backward, and then forward again to acquire the phase difference of $2n\pi + 2\pi$, as required for exiting to A.

As a result, the electrons emitted by B linger in the ring three times longer than the electrons from A, $\tau_{B \to A} = 3\tau_{A \to B}$, which is what Schrödinger's equation says [10] (and I do not feel thoroughly duped up to now).

Next, we introduce inelastic scattering. We consider the case that the mean inelastic scattering time equals $2\tau_{A \to B}$ and that the scattering occurs by a deep trapping site. As the electron becomes trapped, it yields its momentum to the substrate's phonon system. Thereby the electron loses its momentum, phase, and its respective memories. Thermal fluctuations later release the electron from the trap. Having no information on the original travel direction, the electron travels with nominally equal probability to A or to B. The decoherence associated with trapping prevents the electron motion from being controlled by interference with the original part of the wave function on the opposite arm of the ring (me still feeling undeluded).

Now let us put everything together and place such a trapping site into an arm of the ring (Fig. 1). As $\tau_{B \to A} = 3\,\tau_{A \to B}$, the trapping site will catch three times more B➡A than A➡B electrons. But each of the trapped electrons is reemitted to reach A or B with equal probability! The trapping site therefore creates an imbalance by sending back a disproportionately large number of electrons arriving from B.

As a result, these rings let electrons pass preferably in one direction, namely from A to B. If fed by a thermal source in thermal equilibrium, devices of this kind create an imbalance of the electron density between the two contacts. This difference in the electrochemical potential can be used to drive a current through an ohmic resistor, charge a capacitor, or perform work [11]. As my caveat still applies, however, let us ask nature—much as Ted would do—directly in the lab how she has solved this problem.

Happy Birthday, Ted!

**Acknowledgements**

The author gratefully acknowledges the valuable contributions of his collaborators D. Braak, P. Bredol and H. Boschker, as well as helpful discussions with numerous colleagues, in particular T. Kopp.

11. One might pose numerous counterarguments against the operation of the device, such as:

    *Maxwell's demons have been proved not to exist.* The latest case for the non-existence of demons is the Landauer–Bennett erasure argument [12,13]. This argument refers to sentient demons that erase information at an entropy cost, which is not the case here.

    *The second law follows from quantum mechanics.* Entropy does not increase in a unitary evolution of a state. The second law may only be derived from quantum

mechanics with a theory of the measurement process, for which no agreement exists.

*During operation, the device loses entropy to the phonon system of the substrate.* At the start of the experiment, the device is in thermal equilibrium at temperature *T*, *i.e.*, the substrate already has maximal entropy corresponding to *T*.

*Nonreciprocal transport is well known, for example from diodes and quantum rings. It has been shown to be consistent with the second law.* None of these devices operate in linear response.

*The generated voltage cannot be used.* Consider an analogous neutron-sorting device that creates an imbalance of neutron density in A and B.

*Such behavior would already be known.* The approach to describe quantum transport by the transmission of wave packets that become inelastically scattered by specific events to create new quantum states of which the unitary evolution is then calculated, has, to our knowledge, not yet been explored. Nonreciprocal transport across quantum devices has been observed many times experimentally, and induced voltages attributed to artifacts have been measured [14].

12. C.H. Bennett, Stud. Hist. Phil. Mod. Phys. **34**, 501 (2003).

13. H.S. Leff and A.F. Rex, "Maxwell's Demon 2", Institute of Physics Publishing, Bristol and Philadelphia (2003).

14. R. Leturcq, R. Bianchetti, G. Götz, T. Ihn, K. Ensslin, D.C. Driscoll, and A.C. Gossard, Physica E **35**, 327 (2006).

**Figure Caption**

**Figure 1**

Device layout. An *n*-type semiconducting ring is penetrated by a magnetic flux $\Phi$. An electron wave packet that enters the ring is split into two halves that travel along opposite arms of the rings. The magnetic flux and the geometric asymmetry of the device cause a shift of phase difference between these two halves of the wavepacket as it travels around the ring. For electrons moving from A to B, the shift is $2n\pi$, whereas for electrons travelling from B to A it is $(2n+1)\pi$. The red dot *S* marks an inelastic scattering center.

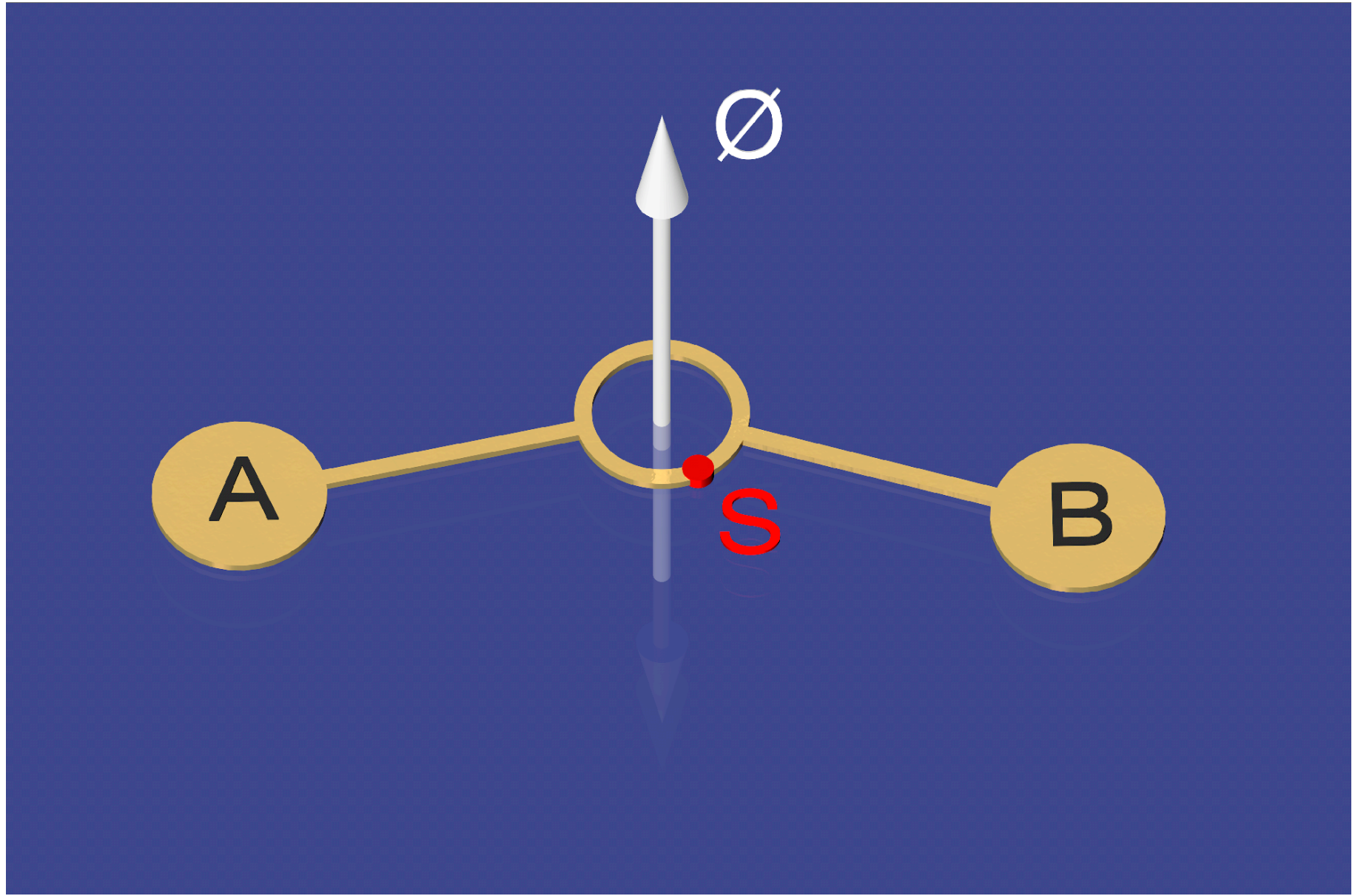